**Symmetry-engineering ferroelectricity in silicon dioxides**

Yin Dai, Menghao Wu*

School of Physics, Huazhong University of Science and Technology, Wuhan 430074, China

It is a long-established rule for classical ferroelectricity that any ferroelectric crystal must adopt one of the 10 specific polar point groups. Here we predict a unique type of ferroelectricity that can be generated in some crystals belonging to nonpolar noncentrosymmetric groups. This principle can be applicable to many systems including silicon dioxides, the most widely used dielectric materials. Most of their crystalline phases do not belong the polar groups, while the nonlinear Si-O-Si configurations lead to multiple identical states. We show first-principles evidence that the crystal symmetry forbidding the formation of polarizations, can be broken by either parallel surfaces in thin-films or applying a uniaxial strain. As a result, the multiple identical states are endowed with polarizations of different directions, and low-barrier ferroelectric switching can be realized via transition between them, which can be room-temperature robust down to the thickness of 1 nm. Our findings may not only enable low-cost and large-scale manufacture of ferroelectrics directly integrated in silicon chips, but also open a new avenue for exploring ferroelectricity in prevalent nonpolar materials.

*Contact author: wmh1987@hust.edu.cn

## I. Introduction

Ferroelectricity, which has been exploited for various applications in the past century, is supposed to exist in crystals only belong to the 10 specific polar point groups (1 ($C_1$), 2 ($C_2$), m ($C_{1h}$), mm2 ($C_{2v}$), 4 ($C_4$), 4mm ($C_{4v}$), 3 ($C_3$), 3m ($C_{3v}$), 6 ($C_6$) and 6mm ($C_{6v}$)). Such basic rule of ferroelectricity can be deduced by Neumann's principle, and the absence of polarization in 11 centrosymmetric and 11 noncentrosymmetric point groups can also be derived. To induce ferroelectricity in prevalent non-polar materials, two symmetry-engineering approaches have been developed respectively for designs of organic and two-dimensional ferroelectrics. For example, functionalization of polar ligands may induce symmetry breaking[1] in non-polar organic crystals. For prevalent two-dimensional materials (e.g. graphene, BN, $MoS_2$) with non-polar crystal lattices, vertical polarizations can be formed upon asymmetric stacking of bilayers/multilayers and switchable via interlayer sliding,[2,3] which have been extensively confirmed by experiments[4-16]. However, these approaches are not applicable to the prevalent dielectric materials widely used in silicon-based circuits.

The incompatibility of prevalent perovskite ferroelectrics with silicon has greatly impeded its potential application in nanoelectronics, since many of the most advanced processing tools (e.g., lithography) are developed specifically for the silicon industry. In recent years, $HfO_2$–based ferroelectrics has emerged as promising materials as they enable low-temperature synthesis and conformal growth on silicon,[17,18] despite their ferroelectricity stem from the metastable asymmetric orthorhombic phase stabilized by surface energy effects and element doping, and similarly for $TiO_2$ thin-film ferroelectrics[19]. Undoubtedly, silicon dioxides are still the most widely used dielectric materials within silicon semiconductor device technology, which can be produced by thermal oxidation of silicon or by using a wide range of vacuum-based techniques, almost ideal with excellent dielectric strength, high resistivity and low defect density. They are not known to be ferroelectrics since their crystals do not belong to the 10 specific polar point groups.

Now this long-established principle is challenged by recent explorations of quantized[20-23] or fractional quantum ferroelectricity[24,25] that may exist in non-polar crystals, where the switching of their quantized polarizations involve long ion displacements of multiple or fractional lattice constants. In 2024, we proposed that the surfaces of crystals are not taken into consideration by Neumann's principle, while the ferroelectricity in such non-polar point groups can be attributed to the breaking of crystal symmetry by the boundaries, which does not decay with increasing size.[26] Similar viewpoint on boundaries has been further promoted in recent studies by Su-huai Wei et al. focused on the nonlocal properties of such ferroelectricity.[27] For example, two-dimensional (2D) binary honeycomb lattices are non-polar with $D_3$ symmetry, while its nanoribbons are always polar as two parallel edges break the 3-fold rotational symmetry and generate inequivalence between 3 directions (as well as two edges) , as illustrated in Fig. 1(a). Similarly, zinc-blende (ZB) structures with space group F-43m belong to the 11 noncentrosymmetric nonpolar groups, while the two parallel surfaces of their thin-films will break the $T_d$ symmetry of the crystal and give rise to spontaneous polarizations. However, the long ion displacements required for switching (e.g., 1/3 lattice constant for the case of 2D honeycomb lattice shown in Fig. 1(b)) usually result in high barriers and current leakage akin to ion conductors, so the potential applications are greatly limited.

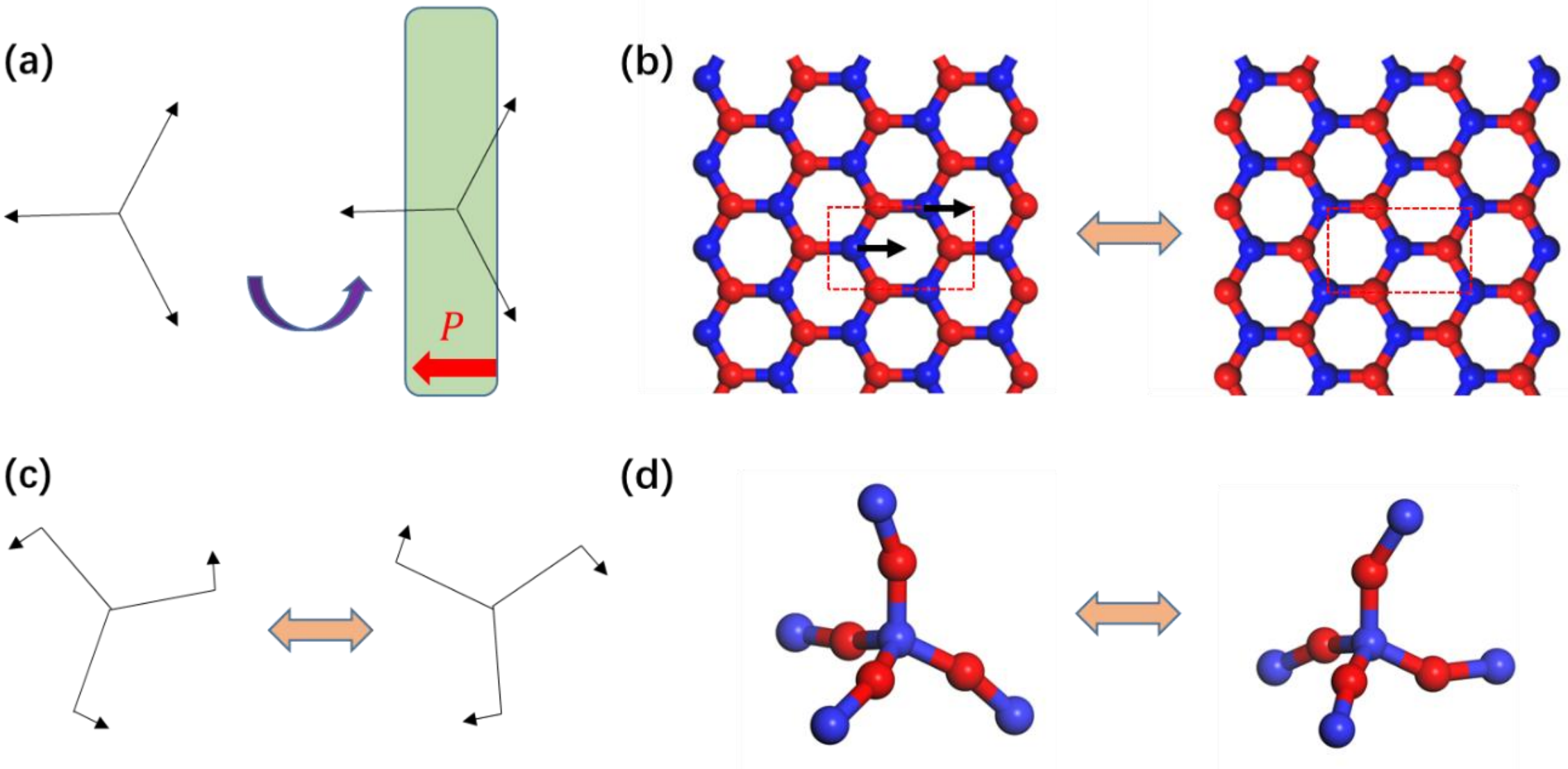

FIG. 1 (a) The $C_3$ symmetry of a crystal will be broken in its ribbon with two parallel edges, giving rise to a polarization. (b) Switching of fractional quantum ferroelectricity in 2D honeycomb lattice via ion displacements of 1/3 lattice constant (marked by the black arrows). (c) 3-fold rotational symmetry with small deviation from $D_3$ may lead to two identical states that are interconvertible via small ion displacements. (d) The distortion induced by nonlinear Si-O-Si configuration in silica may favor the formation of multiple identical states.

In this paper, we propose a possible approach of generating low-barrier ferroelectricity via small ion displacements in non-polar crystals, distinct from integer or factional quantum ferroelectricity. A case in point is the crystal with 3-fold rotational symmetry but small deviation from $D_3$ illustrated in Fig. 1(c), where the two identical states can be interconvertible via small ion displacements. When the 3-fold rotational symmetry is broken by two parallel surfaces, or a uniaxial strain, the two ground states become polar and the transition between them may be electrically driven. Such strategy can be applicable to many crystals that belong to the 11 noncentrosymmetric point groups, despite it fails in centrosymmetric systems where $C_2$ symmetry may sustain upon parallel surfaces or strain. For typical silica crystals constructed by $SiO_4$ tetrahedrons, they are non-centrosymmetric but non-polar. Meanwhile the Si-O-Si bondings are not linear, providing the required distortion for the formation of multiple identical states (see Fig. 1(d)).

## II. Results and Discussion

There are many crystalline allotropes of $SiO_2$, and first we select prevalent α-quartz and β-cristobalite phase (predicted to be the ground state, 33.43 meV/f.u. lower in energy compared with α-quartz phase in previous calculations[28]) as two paradigmatic cases. As the most prevalent crystalline phase of $SiO_2$, α-quartz crystal belongs to a chiral trigonal space group $P3_121$ (right-handed) or $P3_221$ (left-handed), where a three-fold screw axis along the **c** axis leads to a helical structure (**a**, **b**, **c** are its unit vectors). The two structures of α-quartz displayed in Fig. 2(a) are identical ground states correlated by the operation $C_2$ along the z axis, i. e., (x, y, z)→(-x, -y, z). Transition between them can be realized via rotation of $SiO_4$ tetrahedrons, while the barrier in the calculated pathway is only 22 meV/f.u.. However, both states are non-polar so this transition cannot be electrically driven. It is the similar case for β-cristobalite also with multiple non-polar ground states. As shown in Fig. 2(b), its bi-stable states are correlated by mirror symmetry $M_{xy}$ plus translation, which can be interconvertible also via rotation of $SiO_4$ tetrahedrons, as revealed by the low transition barrier in the switching pathway.

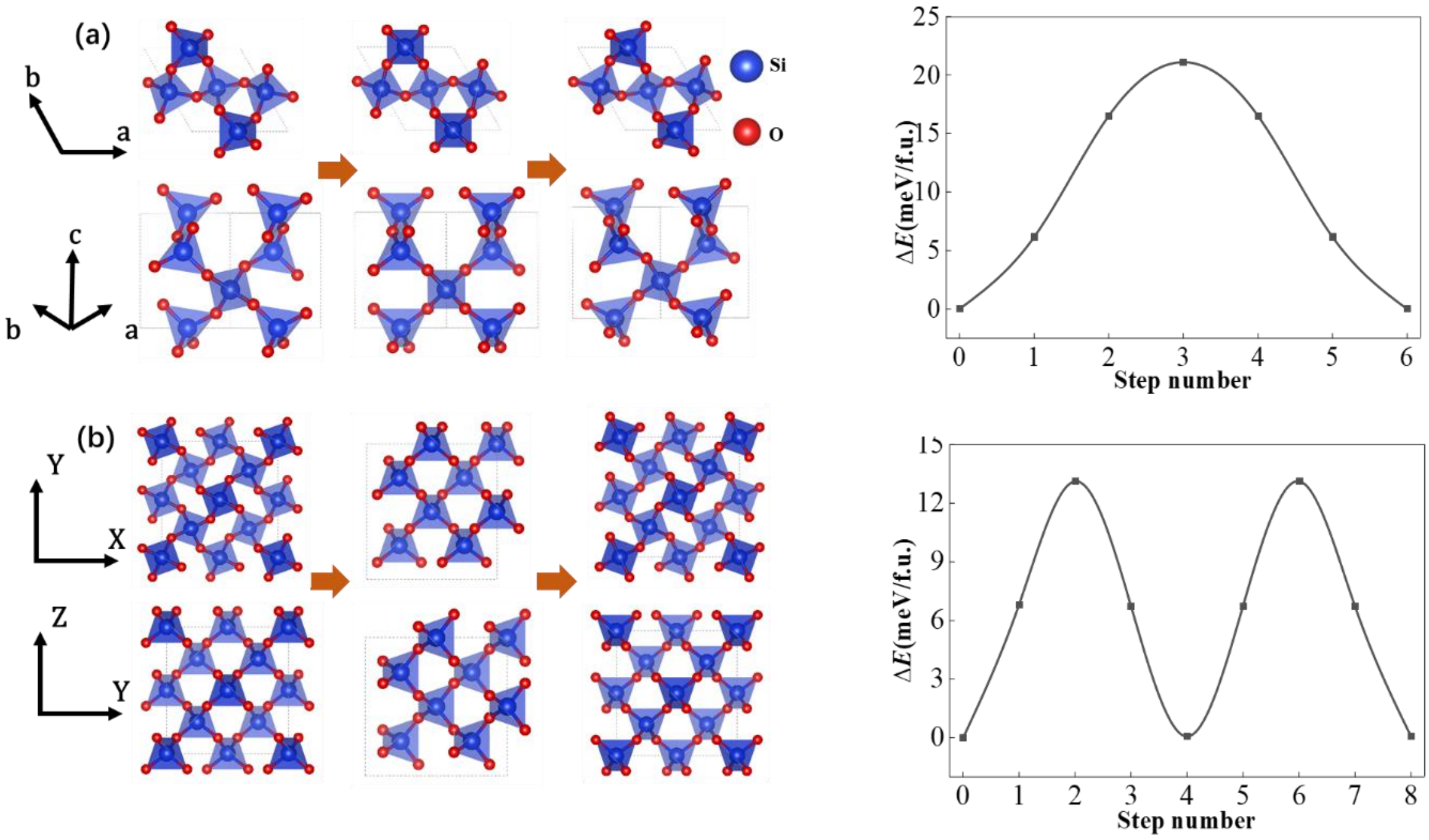

FIG. 2 The transition pathways between two identical nonpolar states for bulk (a) α-quartz and (b) β-cristobalite, where the structural evolutiona of overview and sideview are displayed.

The crystal of α-quartz possesses $C_3$ rotation axis along **c** and $C_2$ rotation axis along **a** and **b**, respectively forbidding the formation of polarization in xy plane and along z axis. When it is cut into slices parallel to the z axis, however, vertical polarization can be induced as the two surfaces break the $C_3$ symmetry. Herein the two identical non-polar states in Fig. 2 may become polar with opposite polarizations, and ferroelectric switching can be realized via transition between them. As shown in Fig. 3(a), the [010] thin-films with thickness around 7-17 Å possess vertical polarizations around 1.1-1.4 pC/m, switchable via the pathway of tetrahedron rotation displayed in Fig. 2. It seems that its ferroelectricity can be robust against depolarization field even below the thickness of 1 nm. Since the crystal structure of α-quartz is known to be highly stable, such ferroelectricity should also be room-temperature robust, which can be revealed by our molecular dynamics (MD) simulations in Fig. S1. Similarly, the polarization in β-cristobalite is forbidden by the $S_4$ symmetry, which may emerge in the thin-film as such symmetry is broken by its two parallel surfaces. As displayed in Fig. 3(b), for [101](or identically [011]) thin-films with thickness 10-20 Å, the estimated vertical polarizations range from 2.6 to 3.1 pC/m, already much higher compared with most 2D sliding ferroelectrics.[2-9] Such polarizations are not quantized, where the switching is realized by the rotation of tetrahedrons instead of long ion displacements for integer/fractional lattice constants.

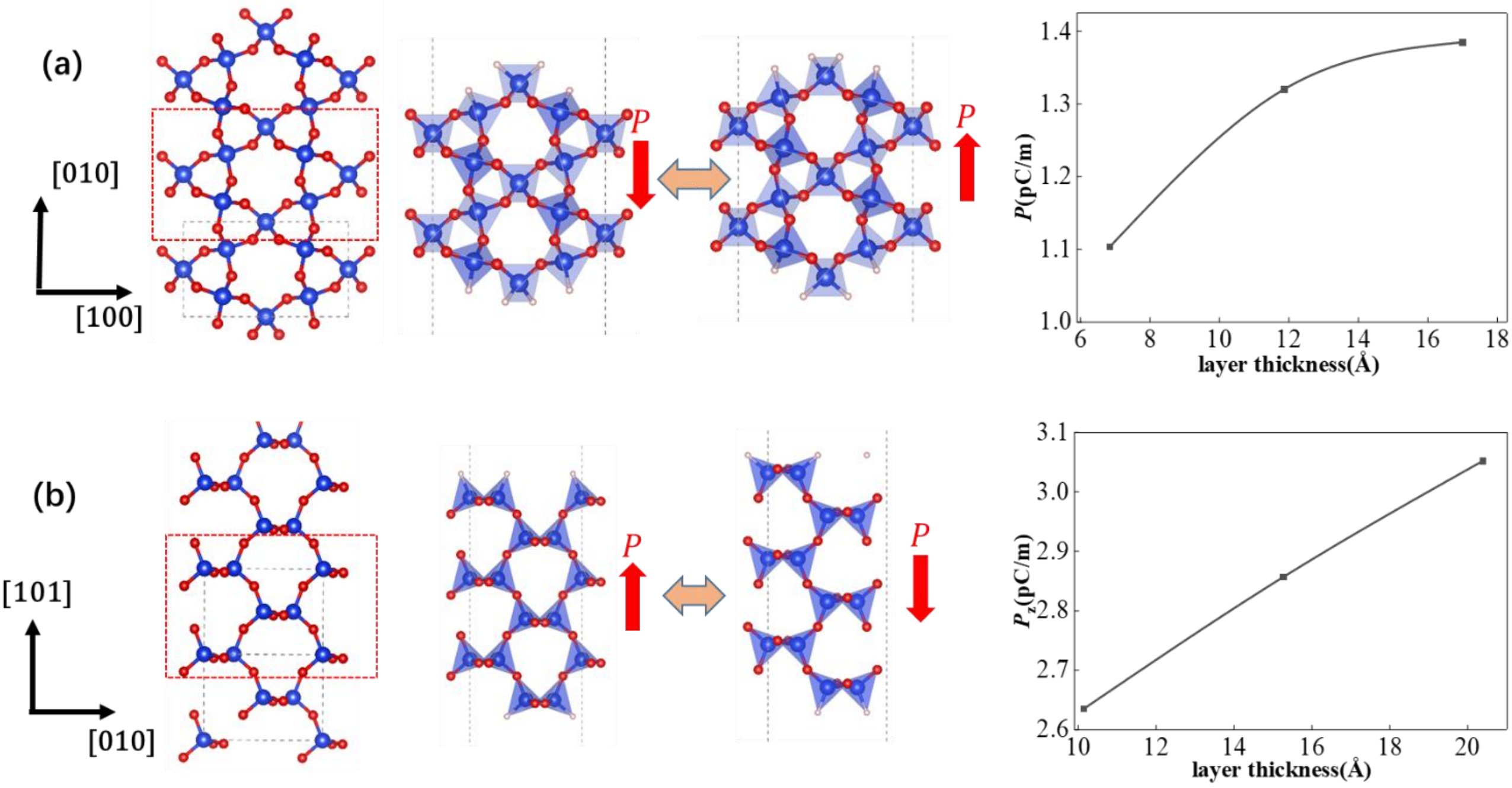

FIG. 3 The formation of vertical ferroelectricity in thin films of (a) [010] α-quartz and (b) [101] β-cristobalite, and the corresponding dependence of polarization on layer thickness.

Strain can be another efficient tool for breaking the crystal symmetry that forbids the formation of polarization. Compared with the symmetry breaking via the surfaces of thin-films, such strain-induced ferroelectricity may exist even in bulk phase. For example, the $C_3$ rotation symmetry of α-quartz can be broken by either an epitaxial or compressive strain along –x direction, as shown in Fig. 4(a). Similarly, the 3 identical directions also become inequivalent if the strain is applied along –y direction. Since the uniaxial strain along either direction cannot break the $C_2$ symmetry along –x direction, the polarization is always formed along –y direction. According to our calculations, the bulk polarization can be enhanced above 0.3 $\mu C/cm^2$ upon a strain of 4% or -1% along –x direction, which can be also achieved by a strain of 3% or -2% along –y direction. Similarly, the $S_4$ symmetry of bulk crystal β-cristobalite, can be broken by uniaxial strain along either -x or –y direction that are equivalent, although the $C_2$ symmetry along two diagonal lines of the unitcell in xy plane will still remain.

As a result, the polarization will be formed along –z direction, which will reach above 0.3 μC/cm$^2$ upon a strain of 2.5% or -2% along either of two directions, as revealed by the strain dependence in Fig. 4(b).

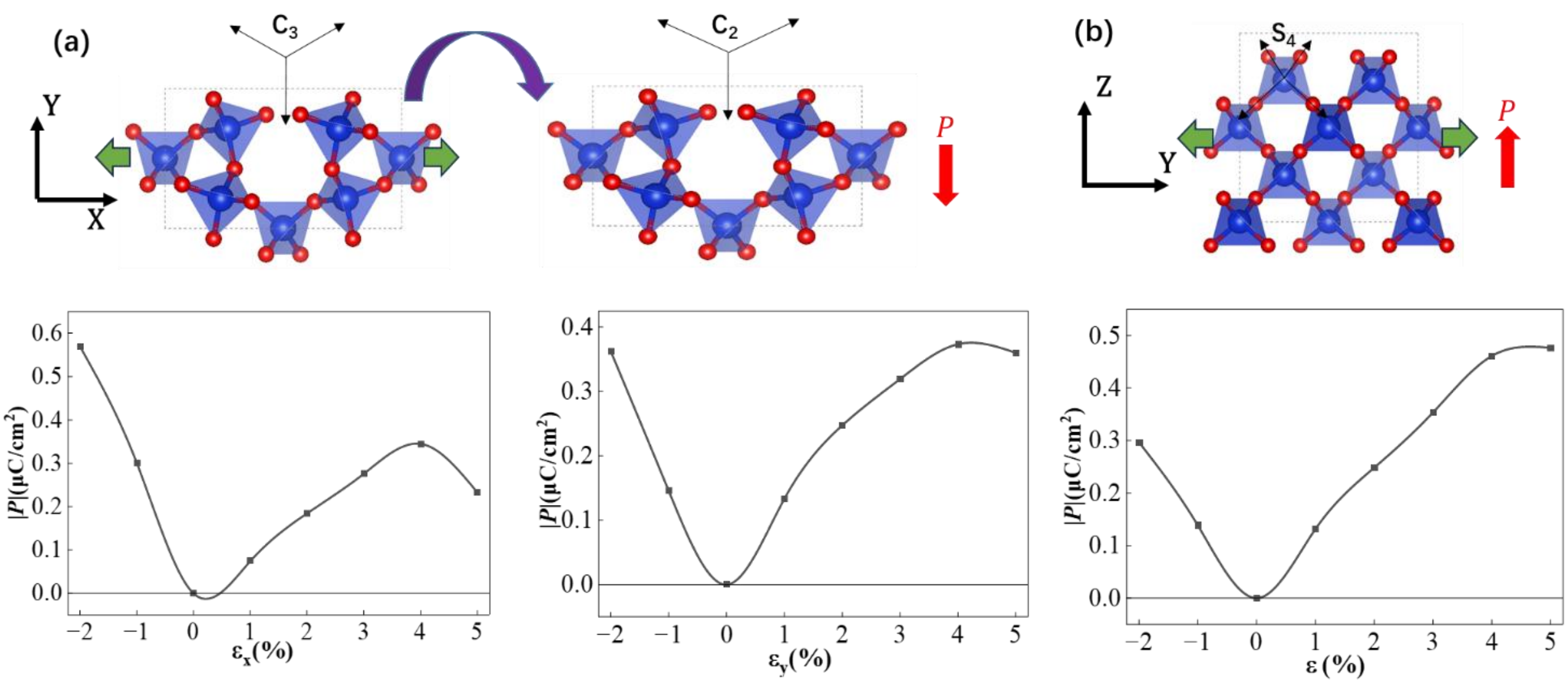

FIG. 4 Dependence of polarization on strain along –x or –y direction for bulk (a) α-quartz and (b) β-cristobalite, where the uniaxial strain can respectively break the $C_3$ and $S_4$ symmetry of two crystals.

It is known that silicon dioxides are usually amorphous on the surface of silicon, which may sometimes be deemed as mixing of various phases. In particular, some metastable phases that belong to the 10 polar groups are even intrinsically ferroelectric, e.g., β-tridymite with a spontaneous polarization of 0.12 μC/cm$^2$, is only 4 meV/f.u. higher in energy compared with the ground state β-cristobalite according to our calculations. As shown in Fig. 5, the ferroelectric switching pathway reveals a switching barrier around 30 meV/f.u.. If the amorphous phase is deemed to be composed of domains of various phases, unconventional ferroelectricity may also be induced by the boundaries and strain at interfaces. Therefore, it will not be astonishing if ferroelectric-like hysteresis is observed in amorphous silicon dioxides on silicon someday.

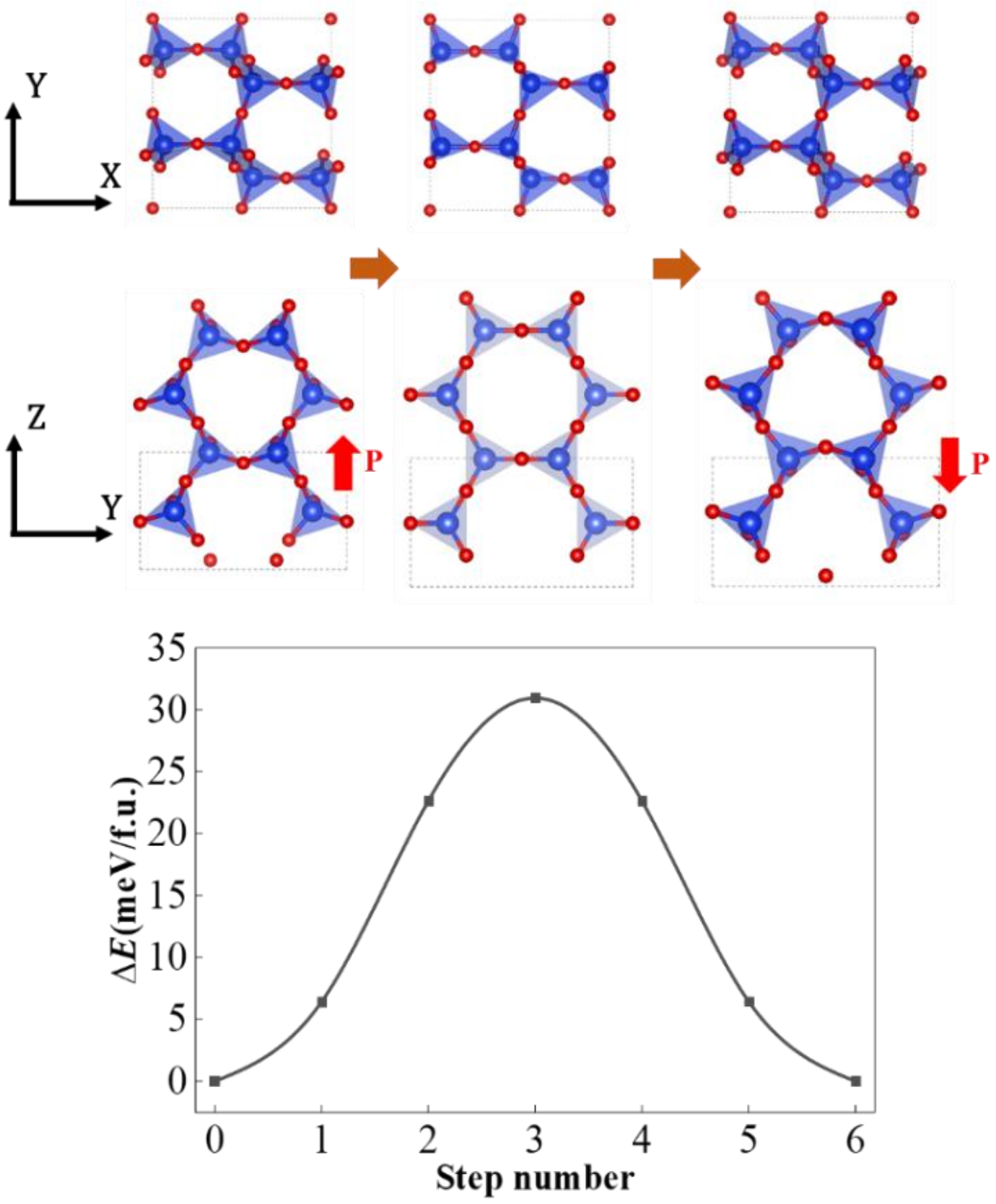

FIG. 5 Ferroelectric switching pathway of β-tridymite.

## III. Conclusions

In conclusion, we propose a strategy of inducing ferroelectricity in crystals that belong to nonpolar noncentrosymmetric groups, and a paradigmatic case is $SiO_2$ with nonpolar crystal structures and nonlinear Si-O-Si configurations that result in multiple identical states. We show that the forbidden polarizations can emerge when the crystal symmetry (e.g., $C_3$ for α-quartz and $S_4$ for β-cristobalite) is broken by either parallel surfaces in thin-films or applying a uniaxial strain. They are switchable via rotation of $SiO_4$ tetrahedrons that leads to the transition between identical states, distinct from quantized ferroelectricity involving ion displacements of multiple/fractional lattice constants. As a result, the bottleneck issue for the integration of ferroelectrics in silicon can be resolved, and their large-scale applications in chips can be expectable. Moreover, this unveiled new mechanism may greatly broaden the scope of ferroelectricity, which should provide vast opportunities of transforming ubiquitous non-ferroelectrics into ferroelectrics.

## IV. Methods

Our calculations are performed within the framework of density functional theory (DFT) implemented in the Vienna Ab initio Simulation Package (VASP 5.4) code[29,30]. The projector augmented wave[31] potential for the core and the generalized gradient approximation in the Perdew-Burke-Ernzerhof[32] form for the exchange correlation functional are applied. The Monkhorst-Pack k-meshes[33] are set to 5 × 7 × 7 in the Brillouin zone for α-quartz, 5 × 5 × 5 for β-cristobalite and 9 × 9 × 5 for β-tridymite, while the electron wave function is expanded on a plane-wave basis set with a cutoff energy of 520 eV. All atoms are relaxed in each optimization cycle until atomic forces on each atom are less than 0.01 $eV·Å^{-1}$, and the energy variation between subsequent iterations falls below $10^{-6}$ eV. The Berry phase method[34] is adopted in computing the ferroelectric polarizations, and a generalized solid-state nudged elastic band method[35] is used to calculate the switching pathway.

## Acknowledgements

The work was supported by National Natural Science Foundation of China (Grant No. 12574263).